\RequirePackage{lineno}
\documentclass[preprint,
eqsecnum,
amsmath,amssymb,
amsmath,amssymb, aps
]{revtex4}

\usepackage[utf8]{inputenc}
\usepackage{bbm} 
\usepackage{indentfirst}
\usepackage{amsfonts}
\usepackage{bm} 
\usepackage{graphicx}
\usepackage{enumitem}
\usepackage{xcolor}
\usepackage{verbatim}

\newcommand{\be}{\begin{equation}}
\newcommand{\ee}{\end{equation}} 
\newcommand{\bea}{\begin{eqnarray}} 
\newcommand{\eea}{\end{eqnarray}}

\newcommand{\A}{\mathbb{A}}
\newcommand{\T}{\mathrm{T}}
\newcommand{\J}{\mathbb{J}}

\begin{document}

\title{Instantons and Fluctuations in a Lagrangian Model of Turbulence}

\author{G. B. Apolin\'ario$^1$, L. Moriconi$^1$ and R. M. Pereira$^2$}
\affiliation{$^1$Instituto de F\'\i sica, Universidade Federal do Rio de Janeiro, \\
C.P. 68528, CEP: 21945-970, Rio de Janeiro, RJ, Brazil}
\affiliation{$^2$Laborat\'orio de F\'isica Te\'orica e Computacional, Departamento de F\'isica, Universidade Federal de Pernambuco, 50670-901, Recife, PE, Brazil}


\begin{abstract}
We perform a detailed analytical study of the Recent Fluid Deformation (RFD)
model for the onset of 
Lagrangian intermittency, within the context of the Martin-Siggia-Rose-Janssen-de Dominicis
(MSRJD) path integral formalism. 
The model is based, as a key point, upon local closures for the pressure Hessian and the viscous 
dissipation terms in the stochastic dynamical equations for the velocity gradient tensor. We carry 
out a power counting hierarchical classification of the several perturbative contributions associated 
to fluctuations around the instanton-evaluated MSRJD action, along the lines of the cumulant expansion. 
The most relevant Feynman diagrams are then integrated out into the renormalized effective action, for 
the computation of velocity gradient probability distribution functions (vgPDFs). While the subleading 
perturbative corrections do not affect the global shape of the vgPDFs in an appreciable qualitative way, 
it turns out that they have a significant role in the accurate description of their non-Gaussian cores.

\end{abstract}

\maketitle

\section{Introduction} \label{sec:Intro}

The non-Gaussian statistical behavior of Galilean-invariant turbulent observables, like inertial 
range velocity differences or velocity gradients -- the hallmark of intermittency \cite{BatchelorTownsend1949} -- is a long 
standing theoretical problem, ultimately related to the existence of nonlocal interactions between eddies 
defined at well-separated spacetime scales across the turbulent energy cascade \cite{TsinoberInformal}.

Fat-tailed velocity gradient probability distribution functions (vgPDFs) are objects of particular
interest in the statistical theory of turbulence \cite{Frisch,Chev2006,ChevPRL}. A main source of motivation has been provided with the introduction, since the mid-1990s, of improved experimental techniques for the
measurement of all of the velocity gradient tensor components, based on specific designs of nine and
twelve-sensor hot wire probes \cite{Wallace9,Wallace2009,WallaceVukoslav2010,KatzSheng2010,tsinober1992experimental,Zeff2003}. It is interesting to remark that the vgPDFs obtained from
these experiments can have their Reynolds number dependence accurately modeled by the phenomenological log-Poisson cascade picture of intermittency \cite{MoriconiPereiraKholmyansky}, which was originally addressed in 
the context of velocity structure functions \cite{she1994universal, dubrulle1994intermittency}. 

Taking into account the fact that the Navier-Stokes equations are nonlocal and non-linear in physical space,
and that at high Reynold's numbers they lead to strong coupling regimes, the Lagrangian picture of the flow 
comes into play as a promising stage for a deeper understanding of intermittency. From a purely 
mathematical perspective, the Lagrangian viewpoint is a natural approach in the context 
of dynamical systems \cite{HolmesLumley}, while it addresses, from its phenomenological side, 
the decoupling of small scale fluctuations from their host large eddies, which break Galilean invariance.
It is clear, however, that effective Lagrangian models are usually formulated at the expenses 
of closure assumptions (which are arbitrary, to some extent) for the dynamics of the velocity 
gradient tensor. 

The simplest of all of the closed Lagrangian models is given by the
Restricted Euler Equation, a model where 
viscous dissipation is neglected and the role of the pressure Hessian is
taken by a velocity gradient-dependent 
term proportional to the identity tensor \cite{leorat1975turbulence,Vieillefosse1982,Vieillefosse1984,MeneveauReview}.
This model yields suggestive results on the classification of turbulent regions, based on the
velocity gradient invariants $Q$ and $R$ \cite{TsinoberInformal}, 
but is, unfortunately, affected by finite-time singularities. Ranging from linear damping to 
stochastic and geometric models \cite{Martin98,Girimaji90,Chertkov99,Jeong2003}, much has been done to solve the instability 
problem, which is, in fact, the main challenging issue to be faced by competing Lagrangian models. A consistent proposal, our focus in this work, is put forward by the Recent Fluid Deformation (RFD) model of Lagrangian turbulence, where dominant contributions to the pressure Hessian and the viscous term are modeled from the strained local evolution, within dissipative time scales, of advected smooth velocity gradient fields \cite{ChevPRL}. 

As it has been recently discussed \cite{MPG}, the RFD model can be recast in the Martin-Siggia-Rose-Janssen-de Dominicis (MSRJD)
path-integral setting \cite{MSRJD1,MSRJD2,MSRJD3}, so that it can be studied with the whole machinery of statistical field theoretical tools. It has been suggested, in this way, that noise renormalization, 
self-induced by non-linear interactions, is the main physical mechanism one needs to consider
in order to understand the onset of fat-tailed vgPDFs, as the Reynold's number increases \cite{MPG},
a point further supported by numerical refinements of the original approach \cite{grigorio2017instantons}.

It is remarkable that despite the use of bold simplifying hypotheses in Ref. \cite{MPG}, which rely essentially on plausibility arguments, the comparison between analytical and empirical vgPDFs turns out to be very convincing. One of the simplifications consists in the use of instantons obtained from a linear truncation of the Euler-Lagrange equations; another one is the selection of a particular form for the contribution of fluctuations around the approximate saddle-point MSRJD action, which just renormalizes the stochastic force-force 
correlation term. It is interesting to call attention, in this connection, to the fact that 
even in the contexts of Eulerian Navier-Stokes (in two or three-dimensions) and Burgers turbulence, 
fluctuations around the instantons cannot be neglected at all in the derivation of the PDF tails 
associated to large vorticities or negative-velocity differences \cite{moriconi2004statistics,falkovich2011vorticity, grafke2015relevance}.

Having in mind that general lessons can be learned from the study of
specific Lagrangian models of intermittency, it is of
great importance to revisit the
analytical treatment of the RFD model \cite{MPG}, in order to
consolidate it as a theoretical
benchmark. This is precisely our aim in this work, which clarifies
what has been missed in the 
previous discussions and also brings to light further improvements
in the description of vgPDFs.
We compare the results of our analytical approach to the ones derived from numerical simulations of the RFD equations, in statistically stationary regimes. These are known to have several similarities with turbulent solutions of the complete Navier-Stokes equations
\cite{ChevPRL}.

This paper is organized as follows. In Sec. \ref{sec:model}, we briefly portray the essential
phenomenological points and the defining equations of the RFD model.
In Sec. \ref{sec:path}, we introduce the general MSRJD path-integral setting for a
large class of tensorial stochastic differential equations, and apply it, in Sec. \ref{sec:Application}, 
to our particular problem of interest. The cumulant expansion is then found to lead to the sum of 
around one hundred perturbative contributions, which are classified in order of importance, on the grounds 
of a power counting analysis. Next, in Sec. \ref{sec:Results}, we carry out a comparison, with the help 
of Monte-Carlo simulations, between the empirical vgPDFs obtained 
from the numerical solution of the RFD stochastic differential equations and the analytical ones derived 
from the field theoretical approach. We also obtain the joint PDFs of the velocity gradient 
invariants $Q$ and $R$ and the local stretching exponents of marginal vgPDFS. Finally, in 
Sec. \ref{sec:Conclusions}, we summarize our results and point out directions of further research.


\section{The RFD Model} \label{sec:model}

The RFD model \cite{ChevPRL} consists of a stochastic differential approach to the onset of Lagrangian intermittency as observed in the turbulent fluctuations of the Lagrangian velocity gradient tensor $ A_{ij}(t) \equiv \partial_j v_i $. It is straightforward to derive, as a starting point, the following integro-differential equation
\begin{equation}  \label{eq:model}
 \dot \A = V[\A] + g \mathbb{F} \ , \ 
\end{equation} 
from the usual stochastic formulation of the Navier-Stokes equations, where $\dot \A \equiv d \A / dt$ ($d/dt$ is the material time derivative) 
and $V[\A]$ is a non-linear and nonlocal functional of  $\A$, defined as
\begin{equation} \label{eq:potential}
 V_{ij} [ \A ] = - (\A^2)_{ij} + \partial_i \partial_j \nabla^{-2} \mathrm{Tr} ( \A^2 ) + \nu \nabla^2 (\A)_{ij} \ . \ 
\end{equation}
The incompressibility condition, $\partial_i v_i = 0$, is equivalent to
$\mathrm{Tr} (\A) = 0$.
In Eq. (\ref{eq:model}), $\mathbb{F} = \mathbb{F}(t)$ is a traceless matrix, whose entries are given by a zero-mean Gaussian stochastic processes with two-point correlators
\begin{equation}
 \langle F_{ij} (t) F_{kl} (t') \rangle \equiv G_{ijkl} \delta (t-t') \ , \ 
\end{equation}
where
\begin{equation}
 G_{ijkl} = 2 \delta_{ik} \delta_{jl} - \frac12 \delta_{il} \delta_{jk} - \frac12 \delta_{ij} \delta_{kl}
\end{equation}
is the most general fourth-order isotropic tensor (up to an overall prefactor) consistent with Eq. (\ref{eq:model}) \cite{Pope2000}.
The stochastic force strength $g$ is proportional to the energy dissipation rate per unit mass, and will play the role of a perturbative coupling constant in our discussions. 
Notice that both $\A$ and $\mathbb{F}$ only depend on time, not on
space, since the RFD models Lagrangian turbulence.

The second and third terms in the right hand side of (\ref{eq:potential}), denoted as the pressure Hessian and the viscous term, respectively, are the focus of modeling in Lagrangian models, where they are replaced by local algebraic functions of the velocity gradient tensor, which, ideally, are designed to preserve important statistical properties derived from their original formulations.

The closure expressions for these two contributions are, in the particular case of the RFD model, obtained from phenomenological arguments related to the short time-scale evolution of small-scale fluid blobs along Lagrangian trajectories.
Since velocity gradients are only correlated on short time-scales,
we may assume that the pressure Hessian and the viscous term are
isotropic at the initial instant of Lagrangian evolution.
In this connection, it is interesting to note that a stronger
assumption -- but a more limited one -- would be to suppose
isotropy at the instant of observation within inviscid dynamics;
this would lead to the Restricted Euler Equation model
\cite{ChevPRL}.

It turns out, in that way, that besides $g$, two time scale parameters,
$\tau$ and $T$, respectively associated to the dissipative and integral
temporal domains, completely define the model, which is assumed to describe
Lagrangian turbulence with Reynolds number $ Re = f(g) (T/\tau)^2 $,
where $f(g)$ is some unknown (probably monotonic) analytical function
of the coupling constant $g$.

The RFD model is, in concrete terms, given by the following approximation to $V [ \A ]$,
\begin{equation}
 V(\A) = - \A^2 + \frac{\mathrm{Tr} (\A^2)}{\mathrm{Tr} (\mathbb{C}^{-1})} \mathbb{C}^{-1} 
 - \frac{\mathrm{Tr} (\mathbb{C}^{-1})}{3 T} \A \ , \
\end{equation}
where $\mathbb{C}$ is the Cauchy-Green tensor,
\begin{equation}
 \mathbb{C} = \exp[ \tau \A] \exp[ \tau \A^\T] \mbox{,}
\end{equation}
which rules the deformation of advected fluid blobs, within dissipative time scales.
We have, thus, from Eq. (\ref{eq:model}),
\begin{equation} \label{eq:dotA}
 \dot \A = V(\A) + g \mathbb{F} = - \A^2 + \frac{\mathrm{Tr} (\A^2)}{\mathrm{Tr} (\mathbb{C}^{-1})} \mathbb{C}^{-1}
 - \frac{\mathrm{Tr} (\mathbb{C}^{-1})}{3 T} \A + g \mathbb{F} \mbox{.}
\end{equation}
Without loss of generality, we take $T=1$ in the above equation. We expand, furthermore, $V(\A)$ up to $O(\tau^2)$, which actually yields a good stage for numerical simulations of the RFD model \cite{afonso2010recent}. To this order, we get
\be \label{vpowers_def}
V(\A) = \sum_{p=1}^4 V_p (\A) \ , \
\ee
where
\begin{subequations} \label{eq:vpowers}
\begin{align}
 V_1 (\A) = &- \A \mbox{,} \\
 V_2 (\A) = &- \A^2 + \frac{\mathbbm{1}}{3} \mathrm{Tr} (\A^2) \mbox{,} \\
 V_3(\A) = &- \frac{\tau}{3} \left( \A + \A^\T - \frac{2 \mathbbm{1}}{3} \mathrm{Tr} (\A) \right) \mathrm{Tr} (\A^2)
 - \frac{\tau^2}{3} \mathrm{Tr} (\A^\T \A) \A  - \frac{\tau^2}{3} \mathrm{Tr} (\A^2) \A \mbox{,} \\
 V_4 (\A) = &- \frac{\mathbbm{1}}{9} \tau^2 \mathrm{Tr} (\A^\T \A) \mathrm{Tr} (\A^2) 
 - \frac{\mathbbm{1}}{9} \tau^2 [ \mathrm{Tr} (\A^2) ]^2
 + \frac{4 \mathbbm{1}}{27} \tau^2 [\mathrm{Tr} (\A)]^2 \mathrm{Tr} (\A^2) \nonumber \\
 &+ \frac{\tau^2}{3} \A^\T \A \ \mathrm{Tr} (\A^2)
 + \frac{\tau^2}{6} (\A^2 + \A^{2 \T}) \ \mathrm{Tr} (\A^2) \mbox{.} 
\end{align}
\end{subequations}
Note that $V_p(\A)$ collects velocity gradient contributions of $O(\A^p)$.
It turns out that the RFD model is able to reproduce,
mostly in a qualitative way, several of the statistical features of
the turbulent fluctuations of the velocity gradient tensor,
as observed either in DNSs or real experiments.
Taking $g=1.0$, the domain of validity of the RFD model,
seen as a ``toy model" of Lagrangian intermittency, is given by the range
$0.05 < \tau < 0.2$, as is empirically suggested from extensive numerical tests 
\cite{ChevPRL}.


\section{Path-Integral Formulation of Stochastic La\-gran\-gian Mo\-dels} \label{sec:path}

Before we concentrate our attention on the particularities of the RFD model, as defined from Eqs.
(\ref{eq:dotA} - \ref{eq:vpowers}), it is interesting to highlight the general path-integral framework 
for the computation of vgPDFs in the setting of closed stochastic Lagrangian models.

Given an arbitrary stochastic differential equation for the velocity gradient tensor $\A$ \cite{comment1},
the MSRJD functional formalism \cite{MSRJD1,MSRJD2,MSRJD3} can be evoked to
express the conditional probability density function of finding $\A = \A_1$ at time $t= 0$, 
provided that $\A = \A_0$ at the initial time $t= -\beta$, as
\begin{equation} \label{eq:rhoPDF}
 \rho ( \A_1 | \A_0, \beta ) \equiv \mathcal{N} \int_{\Sigma} D[\hat \A] D[\A]  \exp \left\{ -S[\hat \A, \A] \right\} \mbox{,}
\end{equation}
where 

\begin{enumerate}[label=(\roman*)]

    \item $\mathcal{N}$ is an unimportant normalization factor (suppressed hereafter in order to simplify notation),
    
    \item the tensor field $\hat \A = \hat \A(t)$ is a time dependent auxiliary tensor field related to the tensorial noise (external stochastic forcing) $\mathbb{F}$, 
    
    \item $\Sigma = \{ \A(-\beta) = \A_0, \ \A(0) = \A_1 \}$ is the set of boundary conditions in the above path-integral,
    
    \item $S[ \hat \A, \A]$ is the so-called MSRJD action, 
    \begin{equation} \label{eq:the_MSRJD_action}
     S[\hat \A, \A] \equiv \int_{- \beta}^0 d t \left\{ i \mathrm{Tr}[\hat \A^\T L(\A) ]
     + \frac{g^2}{2} G_{ijkl} \hat A_{ij} \hat A_{kl} \right\} \mbox{,}
    \end{equation}
    which is a complex-valued quantity, with a mathematical role analogous to the one
    of usual quantum mechanical actions in the Feynman path-integration formalism \cite{feynman2010quantum}
    (as in the above equation, we apply, throughout this paper, Einstein convention to indicate the sum
    over repeated indices in tensor equations). 
    Finally,

    \item $L(\A) \equiv \dot \A - V(\A) $. Note that in the particular case where $\A$ is a solution of (\ref{eq:dotA}), then $L(\A)$ can be identified to the external random forcing  $ g \mathbb{F}$.
    
\end{enumerate}


We are interested in studying vgPDFs for large asymptotic times $\beta \rightarrow \infty$, when it is natural to conjecture that a statistically stationary state for the fluctuations of the velocity gradient tensor has been reached.
Assuming, furthermore, that for asymptotic times the dependence upon the initial condition $\A_0$ has vanished from the conditional vgPDFs, it proves convenient to impose the following periodic boundary condition,
\begin{equation} \label{eq:boundary}
 \A(0) = \A(- \beta) \equiv \bar \A \mbox{,}
\end{equation}
which, as addressed in \cite{MPG}, leads to technical simplifications in the saddle-point approach 
to the MSRJD path-integration.
It is  interesting to note that this choice of periodic boundary conditions is ultimately justified by the fact that the velocity gradient dynamics is correlated on short time scales 
\cite{ChevPRL,chevillard2008modeling}. Accordingly, one may verify that the MSRJD saddle-point action for the RFD model becomes independent on the initial conditions for large evolution times. 
The joint vgPDF for the statistically stationary state
is, therefore,
\begin{equation} \label{asymptPDF}
 \rho(\bar \A) = \lim_{\beta \rightarrow \infty} \rho(\bar \A | \bar \A, \beta) \mbox{.}
\end{equation}

The central idea of our analysis is that the vgPDFs are related to specific dominant flow configurations, 
which are stationary points of the MSRJD action in a functional sense.
In the case of large deviations of fluid dynamical observables -- described by the asymptotic behavior of PDFs' tails --, 
these solutions encode the non-linear/non-local coupling between degrees of freedom
defined at different length scales, as it is clear from
the structure of the Navier-Stokes equations \cite{grafke2015instanton}.
An interesting case study has been provided by the one-dimensional
Burgers equation, where a succesfull convergent numerical scheme
has been implemented for the derivation of instanton solutions
\cite{grafke2015relevance}.

These prevailing configurations can be naturally addressed in the MSRJD path-integration formalism as the functional saddle-points of the MSRJD action -- dubbed as the ``instanton'' fields $\hat \A^{sp}$ and $\A^{sp}$ --, which are derived within the standard steepest descent approach \cite{Amit,ZinnJustin,Peskin} as solutions of the Euler-Lagrange equations,
\begin{equation} \label{eq:EOM}
 \left. \frac{\delta S[\hat \A, \A]}{\delta A_{ij}} \right|_{\substack{\hat \A = \hat \A^{sp} \\ \A = \A^{sp}}} \!\!\!\!\!\! = 0 \ \ \mbox{and} \ \ 
 \left. \frac{\delta S[\hat \A, \A]}{\delta \hat A_{ij}} \right|_{\substack{\hat \A = \hat \A^{sp} \\ \A = \A^{sp}}} \!\!\!\!\!\! = 0  \mbox{,}
\end{equation}
with the periodic boundary condition $ \A^{sp}(0) = \A^{sp}(- \beta) = \bar \A $.

While the instanton fields, solutions of (\ref{eq:EOM}), can be used to devise a first approximation to the vgPDFs, fluctuations around them 
may be in fact relevant in order to achieve reasonable agreement with numerical or experimental results. Denoting the fluctuating fields 
by $\hat \A$ and $\A$, we perform the substitutions $\hat \A \rightarrow \hat \A^{sp} + \hat \A$ and $\A \rightarrow \A^{sp} + \A$ in the MSRJD action, and expand it in
tensor monomials of these fields. The MSRJD action thus takes the form
\begin{equation}
 S[\hat \A, \A] \rightarrow S[\hat \A, \A] = S_{sp}[\hat \A^{sp}, \A^{sp}] + \Delta S[\hat \A, \A] \label{sp_fluct} \mbox{.}
\end{equation}
Note that the above expression is exact and that $ S_{sp}[\hat \A^{sp}, \A^{sp}] $ and $ \Delta S[\hat \A, \A]$ are
the contributions to the MSRJD action that contain, respectively, {\it{only}} the instanton fields, and all
the additional terms that involve the fluctuations $\A$ and $\hat \A$. 
The saddle-point action $S_{sp}[\hat \A^{sp}, \A^{sp}]$ is simply the 
MSRJD action, (\ref{eq:the_MSRJD_action}), evaluated with the instanton
fields, $\hat \A^{sp}$ and $\A^{sp}$.
From (\ref{eq:rhoPDF}), (\ref{asymptPDF}), and (\ref{sp_fluct}), the vgPDF can be correspondingly rewritten as
\begin{equation}
 \rho(\bar \A) = \exp \left\{ -S_{sp}[\hat \A^ {sp}, \A^ {sp}] \right\} \int D[\hat \A] D[\A] \exp \left\{ -\Delta S[\hat \A, \A] \right\} \mbox{.}
\end{equation}
As a well-established procedure in statistical field theory \cite{kogut1979introduction,Barabasi,Kardar}, the above path-integration over fluctuations can be perturbatively computed within the non-interacting model given by $ \Delta S_0 [\hat \A, \A] $,
defined as the quadratic contribution to the MSRJD action that is independent of the saddle-point solutions. We 
just mean that $ \Delta S[\hat \A, \A]$ can be exactly split as
\begin{equation} \label{deltaS}
 \Delta S[\hat \A, \A] = \Delta S_0[\hat \A, \A] + \Delta S_1[\hat \A, \A] \mbox{,}
\end{equation}
where $ \Delta S_1$ contains all the self-interacting terms of the MSRJD action. We obtain
\begin{equation} \label{eq:pdf}
 \rho(\bar \A) = \exp \left\{ -S_{sp}[\hat \A^ {sp}, \A^ {sp}] \right\}
 \left\langle \exp [ -\Delta S_1 ] \right\rangle_0 \ , \ 
\end{equation}
where the expectation value $ \left\langle \exp [ -\Delta S_1 ] \right\rangle_0 $ is computed 
in the model defined by the quadratic action $ \Delta S_0$.
If we write now, up to a normalization factor, the non-normalized vgPDF as
\be
 \rho(\bar \A) = \exp \left\{ -\Gamma[\hat \A^{sp}, \A^{sp} ] \right\} \mbox{,} \label{nn_vgPDF}
\ee
then the cumulant expansion comes into play as a pragmatic way to approximate $\Gamma[\hat \A^{sp}, \A^{sp} ]$
from the evaluation of statistical moments of $\Delta S_1$. Up to second order in $\Delta S_1$, we get
\begin{equation} \label{eq:massive}
\begin{split}
 \Gamma[\hat \A^{sp}, \A^{sp} ] &= 
 S_{sp}[\hat \A^{sp}, \A^{sp} ] + \langle \Delta S_1[\hat \A^{sp}, \A^{sp} ] \rangle_0 \\
 &- \frac{1}{2}
 \left( \langle \Delta S_1^2[\hat \A^{sp}, \A^{sp} ]  \rangle_0 - \langle \Delta S_1[\hat \A^{sp}, \A^{sp} ] \rangle_0^2 \right) \mbox{.}
\end{split}
\end{equation}
The MSRJD effective action $\Gamma[\hat \A^{sp}, \A^{sp} ]$ still satisfies, to first order in the perturbations, the equations of motion (\ref{eq:EOM}). 

It is important to emphasize -- a point of pragmatical relevance -- that it is in general difficult to find exact solutions of the saddle-point Eqs. (\ref{eq:EOM}). This, however, should not be a matter of great concern, if one is able to find reasonable approximations for the instanton fields, since the substitution (\ref{sp_fluct}) and the second order cumulant expansion result (\ref{eq:massive}) are always meaningful perturbative procedures in weak coupling regimes. 

The cumulant expansion terms are represented in general as Feynman diagrams, and, depending on the particular model under study, they can be numerous, with variable relative weights. A power counting procedure would then be suitable to single out the most relevant diagrams.

In the present work, more specifically, we center our attention on the description of 
non-Gaussian velocity fluctuations near the vgPDF cores. In this case, provided that
the Reynolds numbers are not too high, we can approximate the exact instantons by convenient closed analytical expressions which can be derived from the quadratic contributions to the MSRJD
action. It  is important to have in mind that such a simplification
would not work if one would be interested to model the far vgPDF tails, essentially dependent on the exact non-linear instantons (in cases where the vgPDFs' tails decay faster than a simple exponential).


\section{Application to the RFD Model}  \label{sec:Application}

The approach discussed in the previous section can be straightforwardly applied to the set up of the RFD model. Following the notation introduced in (\ref{deltaS}), we have, for the RFD model,
\begin{equation} \label{eq:quadratic}
 \Delta S_0[\hat \A, \A] = \int_{- \beta}^0 d t \left\{ i \mathrm{Tr} [ \hat \A^\T (\dot \A + \A)
 + \frac{g^2}{2} G_{ijkl} \hat A_{ij} \hat A_{kl} \right\}
\end{equation}
and, using (\ref{vpowers_def}),
\begin{equation} \label{eq:deltaS1}
 \begin{split}
  \Delta S_1 [ \hat \A, \A] &= - i \sum_{p=2}^{4} \int_{- \beta}^0 d t \ 
  \mathrm{Tr}[ ( \hat \A^{sp} )^\T ( V_p(\A) + \Delta V_p (\A) ) ] \\
  &+ \mathrm{Tr}
  [ \hat \A^\T V_p (\A^{sp})] + \mathrm{Tr} [ \hat \A^\T ( V_p(\A) + \Delta V_p (\A) ) ] \mbox{,}
 \end{split}
\end{equation}
where 
\begin{equation}
 \Delta V_p(\A) = V_p( \A^{sp} + \A ) - V_p( \A^{sp} ) - V_p( \A ) \ . \
\end{equation}
Note that if the $V_p(\A)$ were linear functions of $\A$, we would have 
$\Delta V_p(\A) = 0$. Furthermore, it is not difficult to see that the tensor 
monomials in the expansion of $\Delta V_p(\A)$ mix the instanton and
the fluctuating fields.

\begin{figure}[ht]
 \centering
 \includegraphics[width=.26\textwidth]{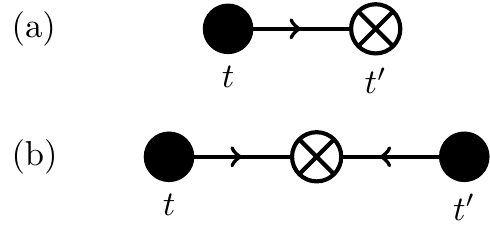}
 \caption{The unperturbed two-point correlation functions of the RFD model, given by diagrams (a) and (b), respectively related to the time translation invariant expressions (\ref{eq:q_propagator}) and (\ref{eq:0_propagator}).}
 \label{fig:free-propagator}
\end{figure}

The two-point correlators (propagators) associated to the tensor fields $\hat \A$ and $\A$ can be calculated through second order functional derivatives of the free-model generating functional,
\begin{equation}
 Z[\J,\hat \J] = \int D[\hat \A] D[\A] \exp \left\{ - \Delta S_0[\hat \A, \A] +i \int_{ - \beta}^0 d t \ \mathrm{Tr} [\hat \J \A + \J \hat \A] \right\} \mbox{,}
\end{equation}
with respect to the external source fields $\hat \J$ and $\J$, at
$\hat \J = \J = 0$. We find, in this way, the causal propagator,
\begin{equation} \label{eq:q_propagator}		
\langle A_{ij} (t) \hat A_{kl} (t') \rangle_0 = \left. \frac{\delta^2 \ln(Z[\J,\hat \J])}{\delta J_{ji}(t) \delta \hat J_{lk}(t')} \right |_{\hat \J = \J=0}= -i \theta (t-t') \exp(t'-t) \delta_{ik} \delta_{jl}
\end{equation}
and, also, the two-point velocity gradient correlation function
\begin{equation} \label{eq:0_propagator}
 \langle A_{ij} (t) A_{kl} (t') \rangle_0 =
 \left. \frac{\delta^2 \ln(Z[\J,\hat \J])}{\delta J_{ji}(t) \delta J_{lk}(t')} \right |_{\hat \J = \J=0} = \frac{g^2}{4} \exp(-|t-t'|) G_{ijkl} \mbox{,}
\end{equation}
which are represented as the Feynman diagrams illustrated in Fig. \ref{fig:free-propagator}.

\begin{figure}[ht]
    \centering
    \includegraphics{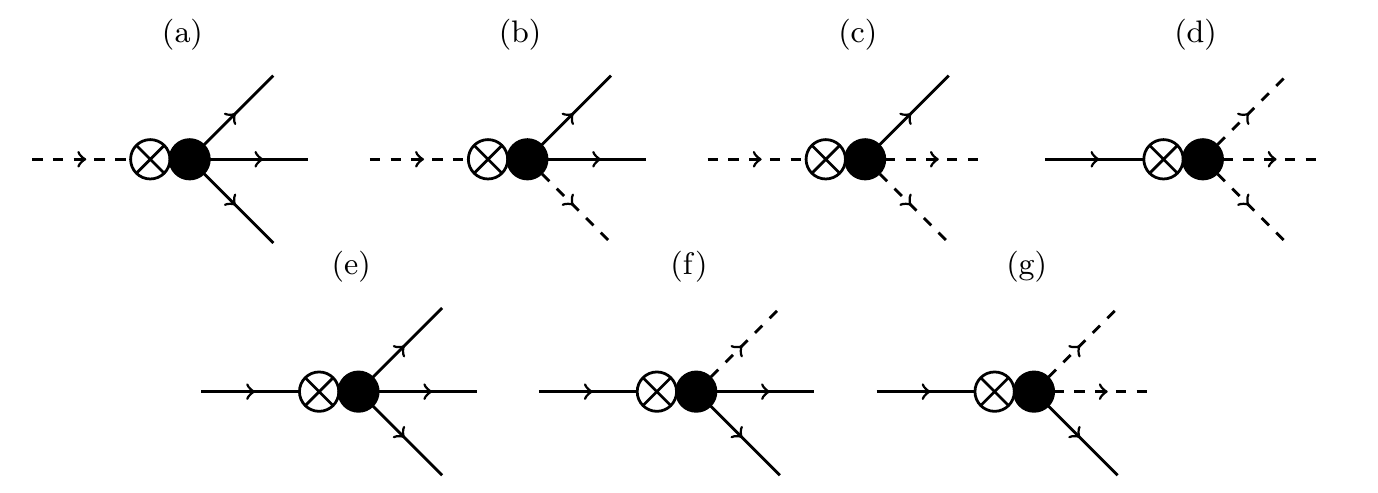}
    \caption{Fourth-order vertices taken from the MSRJD action for the RFD model, Eq. (\ref{eq:deltaS1}). Dashed lines attached to crossed or filled circles, 
    indicate, respectively, the insertion of the instanton fields $ \hat \A^{sp}$ (dashed incoming lines) or
    $\A^{sp}$ (dashed outgoing lines) in the perturbative vertices. Solid lines have an analogous interpretation, given in
    terms of the fluctuating fields $ \hat \A$ and $\A$. These vertices are related to the following contributions to the MSRJD action (by ``odd'' or ``even'' parts of traces, we refer to the sum of tensor monomials that contain an odd or even total number of fluctuating fields): 
    (a) $ \mathrm{Tr} [ (\hat \A^{sp})^\T V_3(\A) ] $,
    (b) odd part of $ \mathrm{Tr} [  ( \hat \A^{sp})^\T \Delta V_3(\A) ]$,
    (c) even part of $ \mathrm{Tr} [  ( \hat \A^{sp})^\T \Delta V_3(\A) ]$,
    (d) $ \mathrm{Tr} [ \hat \A^\T V_3 (\A^{sp})] $,
    (e) $ \mathrm{Tr} [ \hat \A^\T V_3(\A) ] $,
    (f) odd part of $ \mathrm{Tr} [ \hat \A^\T \Delta V_3(\A) ] $ and
    (g) even part of $ \mathrm{Tr} [ \hat \A^\T \Delta V_3(\A) ] $.}
  \label{fig:Vertices}
\end{figure}

The several contributions to the MSRJD action that come from Eq. (\ref{eq:deltaS1}) yield the diagrammatic vertices that are used to build up the perturbative expansion of general correlation functions \cite{Barabasi,Kardar}, from the application of Wick's theorem \cite{Amit,ZinnJustin}. As an example, the complete set of fourth-order vertices is depicted in Fig. \ref{fig:Vertices}.

In connection with (\ref{eq:massive}) and (\ref{eq:deltaS1}), we collect one hundred and eleven Feynman diagrams that should be computed in order to get the effective MSRJD action up to second order in the cumulant expansion. Even though such a time consuming evaluation is likely to be within the reach of present algebraic computational methods, we can show that the vast majority of these contributions can be actually neglected in the perturbative regimes of interest. Our rationale to achieve this simplification is based on a careful determination of the powers of the coupling parameters $g$ and $\tau$, and also of the powers of the instanton fields associated to each one of the Feynman diagrams. 

Usual graph-theoretical arguments, combined with the saddle-point equations given in (\ref{eq:EOM}), which allow us to express $\hat \A^{sp}$ in terms of $\A^{sp}$, imply that any given diagram in the cumulant expansion with $L$ loops, $E$ external lines (representing $\hat \A^{sp}$ or $\A^{sp}$ fields), and $N_3$ and $N_4$ vertices of types $\hat \A\A\A\A$ and $\hat \A\A\A\A\A$  (for $\hat \A$ and $\A$ being saddle-point or fluctuating fields), respectively, is proportional to
\begin{equation} \label{eq:powers}
 g^{2(L-1)} (1 + a\tau^{N_3})\tau^{N_3 + 2 N_4} f( \A^{sp} ) \mbox{,}
\end{equation}
where $f( \A^{sp} )$ is a diagram-dependent homogeneous scalar function of $\A^{sp}$ with homogeneity degree $E$ (that is $f( \alpha \A^{sp}) = \alpha^E f( \A^{sp} )$, for any real positive parameter $\alpha$), and $a$ is an unimportant constant (of the order of unity). It is important to note that vertices of type $ \hat \A \A \A$ do not contribute with factors that depend on their diagrammatic participation number $N_2$, since these diagrams derive from $V_2$ contributions, which do not depend on $\tau$, as it can be seen very clearly from Eq. (\ref{eq:vpowers}b). Thus, for each Feynman diagram that takes part in the cumulant expansion, we define, taking into account (\ref{eq:boundary}) and (\ref{eq:powers}), its ``power counting coefficient'', as
\begin{equation}
C(g,\tau, A) = g^{2(L-1)} {\hbox{Max($\tau^{N_3}$,$\tau^{2N_3}$)}} \tau^{2 N_4} A^E \label{power_count_coeff} \ , \ 
\end{equation}
where 
\begin{equation}
A \equiv \sqrt{ \mathrm{Tr} [ \bar \A^T  \bar \A] }
\end{equation}
is a measure of the velocity gradient strength for the velocity gradient tensor $\bar \A$ where the vgPDF is evaluated. 

\begin{figure}[ht]
 \centering
 \includegraphics[width=\textwidth]{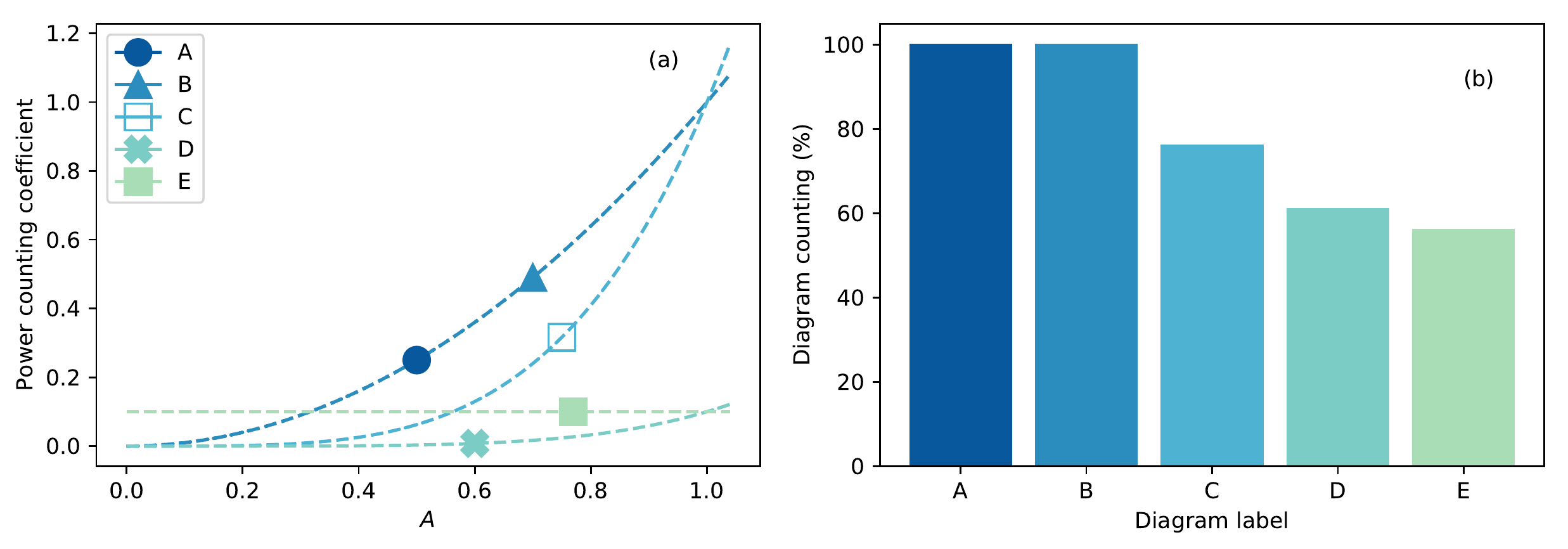}
 \caption{(a)  Plots of the power counting coefficient as a function of the velocity gradient strength $A$, as defined from the expectations values (\ref{eqA}-\ref{eqE}) taken for $g=1.0$ and $\tau=0.1$. (b) Relative frequencies, within the interval $0 \leq A \leq 1$, of the cases where the power counting coefficients are found to be among the first five largest ones.}
 \label{fig:powers-scaling}
\end{figure}

With the help of Eq. (\ref{power_count_coeff}), we establish, then, a rank of relevance for the diagrams of interest, as $A$ is varied for fixed values of $g$ and $\tau$. In consonance with previous numerical studies \cite{ChevPRL,MPG,afonso2010recent,grigorio2017instantons}, we take, for the ranking analysis, $g=1.0$ and $\tau=0.1$. The numerical values of the power counting coefficients are inspected in the interval $0 \leq A \leq 1$,
a range where perturbation theory is assumed to hold,
a fact
we verify \textit{a posteriori} from the computation of vgPDFs.
It is important that $\tau$ be small enough for the RFD model to
yield statistical results which are qualitatively similar to
the ones derived from the exact Navier-Stokes equations.
The forcing parameter $g$ is our main perturbative parameter,
since it controls the intensity of fluctuations.
In addition, the velocity gradient strength $A$ must also be small,
once we are focused on the evaluation of
relevant intermittency corrections near the vgPDF cores. 
The most important five contributions that appear more frequently in that range of velocity gradient strengths are labeled, in ranking order of decreasing importance, with boldface letters from {\bf{A}} to {\bf{E}}, and correspond to the following cumulant expansion terms,
\begin{align}
  &\textbf{A:}  \  \langle \mathrm{Tr} \left [ ( \hat \A^{sp})^\T V_2(\A) \right ]_t  \mathrm{Tr} \left [ ( \hat \A^{sp})^\T V_2(\A) \right ]_{t'}   \rangle_0  \sim A^2 \ , \ \label{eqA} \\ 
  &\textbf{B:} \ \langle \mathrm{Tr} \left [ ( \hat \A^{sp})^\T V_2(\A) \right ]_t  \mathrm{Tr} \left [ \hat \A^\T \Delta V_2(\A) \right ]_{t'} \rangle_0 \sim A^2 \ , \ \label{eqB} \\ 
  &\textbf{C:} \ \langle \mathrm{Tr} \left [ ( \hat \A^{sp})^\T \Delta V_2(\A) \right ]_t  \mathrm{Tr} \left [ \hat \A^\T V_2(\A^{sp}) \right ]_{t'} \rangle_0 \sim  A^4/g^2 \ , \ \\
  &\textbf{D:} \ \langle \mathrm{Tr} \left [ ( \hat \A^{sp})^\T \Delta V_3(\A) \right ]_t \mathrm{Tr} \left [ \hat \A^\T V_2(\A^{sp}) \right ]_{t'} \rangle_0 \sim  \tau A^5/g^2 \label{eqD} \ , \ \\
  &\textbf{E:} \ \langle \mathrm{Tr} \left [ ( \hat \A^{sp} )^\T V_3(\A) +  \hat \A^\T V_3 (\A^{sp}) +  \hat \A^\T V_3(\A) \right ]_t \rangle_0 = {\hbox{const.}}  \label{eqE} \ , \ 
\end{align}
which have their power counting coefficients plotted in Fig. \ref{fig:powers-scaling}a. The histogram analysis of the above top five expectation values is furthermore given in Fig. \ref{fig:powers-scaling}b.
These five diagrams agree with the intuitive notion, brought by
Eq. (\ref{power_count_coeff}), that the more relevant diagrams have smaller values
of $N_3$ and $N_4$, since $\tau = 0.1$. As a matter of fact, we point out that the higher order terms produced from the second order cumulant contributions, and not scrutinized in Eqs. (\ref{eqA} - \ref{eqE}), have prefactors that are proportional to powers of the small parameter $\tau$, and, thus, play a negligible role in the account of perturbative contributions.

It turns out that in the considered range of velocity gradient strengths, two contributions, which have exactly the same power counting coefficients,
are clearly dominant over the remaining ones. These are the cumulant corrections {\hbox{\bf{A}}} and {\hbox{\bf{B}}}, defined in Eqs. (\ref{eqA}) and (\ref{eqB}). 
Note that the power counting coefficient for the contribution {\hbox{\bf{E}}}, Eq. (\ref{eqE}), is actually independent of $A$, and, therefore, plays 
no role at all in the evaluation of vgPDFs.
Diagram \textbf{E} is only displayed for matters of completeness,
since it casually happens to be larger than many other diagrams. It is important to note that power counting is actually an effective way to identify relevant contributions, provided these are in fact dependent on the velocity gradient tensor - a fact that we check for each one of the selected diagrams.

\begin{figure}[htb]
  \centering
  \includegraphics{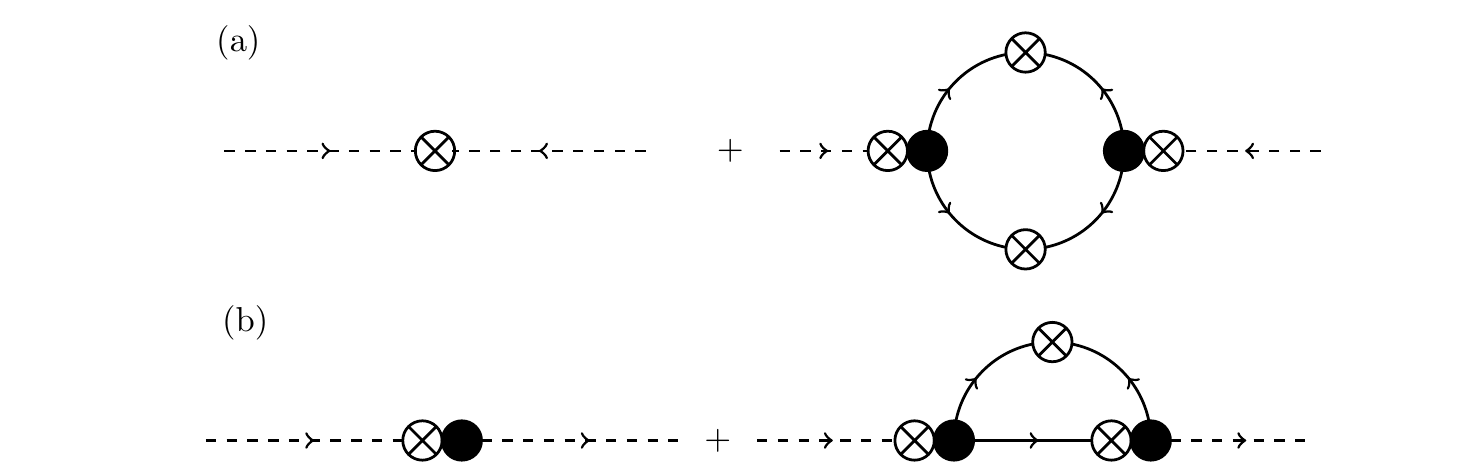}
  \caption{Feynman diagrams for (a) the renormalized noise and (b) the renormalized causal propagator kernels, which take into account the one-loop contributions {\bf{A}} and {\bf{B}}, respectively.}
  \label{fig:oneloop}
\end{figure}

From Eq. (\ref{eq:massive}) and the fact that $ \langle \Delta S_1 [\hat \A, \A] \rangle_0 = 0$, we conclude that the MSRJD effective action can be written
as
\begin{equation} \label{eq:eff_MSRJD_action}
 \Gamma[\hat \A^{sp}, \A^{sp} ] = S[\hat \A^{sp}, \A^{sp} ] + \sum_n C_n [\hat \A^{sp}, \A^{sp} ] \mbox{,}
\end{equation}
where $n$ labels the several second order cumulant expansion terms $C_n [\hat \A^{sp}, \A^{sp} ]$, which in our case are dominated by the contributions $\textbf{A}$ and $\textbf{B}$.
Their associated Feynman diagrams, represented in Fig. \ref{fig:oneloop}, are noted to renormalize the noise and propagator kernels in the effective MSRJD action (\ref{eq:eff_MSRJD_action}).
The contributions {\bf{A}} and {\bf{B}} to the effective action (\ref{eq:eff_MSRJD_action}) can be written, more concretely, as
\begin{equation}\label{eq:diag_a_effective}
 C_{\bf{A}} [\hat \A^{sp}, \A^{sp} ]  = \frac12 \int_{-\beta}^0 d t \int_{-\beta}^0 d t'
 \hat A^{sp}_{ij} (t) \hat A^{sp}_{kl} (t') C^A_{ijkl} (t-t')
\end{equation}
and
\begin{equation}\label{eq:diag_b_effective}
 C_{\bf{B}} [\hat \A^{sp}, \A^{sp} ]  = \frac12 \int_{-\beta}^0 d t \int_{-\beta}^0 d t'
 \hat A^{sp}_{ij} (t) C^B_{ij} (\A^{sp}(t'),t-t') \mbox{,}
\end{equation}
where
\begin{equation}
    C^A_{ijkl} (t-t') = \langle [ V_2 (\A(t)) ]_{ij} [V_2 (\A(t')) ]_{kl} \rangle_0
\end{equation}
and
\begin{equation}
   C^B_{ij} (\A^{sp}(t'),t-t') = \langle [ V_2 (\A(t)) ]_{ij} \hat A_{kl} (t') [\Delta V_2 (\A(t')) ]_{kl} \rangle_0 \mbox{.}
\end{equation}
\vspace{0.3cm}

\subsection{Structure of the MSRJD Effective Action}

The effective action (\ref{eq:eff_MSRJD_action}) can be written, after the introduction of the contributions (\ref{eq:diag_a_effective}) and (\ref{eq:diag_b_effective}), as

\begin{equation}\label{eq:effective_with_integrals}
 \Gamma [ \hat \A, \A] = i \int_{-\beta}^0 d t \int_{-\beta}^0 d t' \left \{ 
  \mathrm{Tr} [ \hat \A^\T(t) L^{\mathrm{ren}}(\A(t'),t-t') ] 
 + \frac{g^2}{2} G_{ijkl}^{\mathrm{ren}} (t-t') \hat A_{ij} (t) \hat A_{kl} (t') \right \}  \mbox{,}
\end{equation}
where
\begin{equation}\label{eq:noise_renormalized}
G_{ijkl}^{\mathrm{ren}} (t-t') \equiv  G_{ijkl} \delta(t-t') + C^A_{ijkl}(t-t') 
\end{equation}
and
\begin{equation}\label{eq:potential_renormalized}
L_{ij}^{\mathrm{ren}}(\A(t'),t-t') \equiv L_{ij}(\A(t')) \delta(t-t') + C^B_{ij}(\A(t'),t-t') \ . \ 
\end{equation}
In contrast to the original nonperturbed MSRJD action (\ref{eq:quadratic}), the above renormalized form (\ref{eq:effective_with_integrals}) contains kernels that depend
non-trivially on a pair of time instants $t$ and $t'$. As it is usual (sometimes in an implicit way) in renormalization group
studies \cite{Amit,ZinnJustin,Peskin,kogut1979introduction,Barabasi,Kardar}, the structure of the renormalized effective action can be simplified in the case of slowly varying fields
(as the instanton fields are assumed to be). This simplification is achieved through the procedure of low-frequency renormalization, which 
in our context consists in replacing the renormalization kernels $C^A_{ijkl}$ and $C^B_{ij}$ by singular ones, 
according to the prescriptions
\begin{subequations}
\begin{align}
&C^A_{ijkl}(t-t') \rightarrow \tilde C^A_{ijkl} \delta(t-t') \ , \ \\
&C^B_{ij}(\A(t'), t-t') \rightarrow \tilde C^B_{ij}(\A(t')) \delta(t-t') \ , \ 
\end{align}
\end{subequations}
where
\begin{subequations}
\begin{align}
& \tilde C^A_{ijkl} \equiv  \int_{- \infty}^\infty dt'  C^A_{ijkl}(t- t') \ , \ \label{eq:coeff_diag_a}\\
& \tilde C^B_{ij}(\A(t)) \equiv \int_{- \infty}^\infty dt'  C^B_{ij}(\A(t'),t-t') \ . \ \label{eq:coeff_diag_b}
\end{align}
\end{subequations}
Substituting (\ref{eq:coeff_diag_a}) and (\ref{eq:coeff_diag_b}) in (\ref{eq:noise_renormalized}) and 
(\ref{eq:potential_renormalized}), the nonperturbed and the effective MSRJD actions will, then, become
isomorphic to each other, provided that the tensors
$G_{ijkl}$ and $L_{ij}(\A)$ of the nonperturbed action are mapped, respectively, to the tensors
\be
\tilde G_{ijkl}^{\mathrm{ren}} \equiv  G_{ijkl} + \tilde C^A_{ijkl} 
\ee
and
\be
\tilde L_{ij}^{\mathrm{ren}}(\A) \equiv L_{ij}(\A)  + \tilde C^B_{ij}(\A) 
\ee
that appear in the definition of the effective renormalized action.

It is important to observe, furthermore, that from the traceless property of the stochastic forcing, it follows that $\tilde G_{iikl}^{\mathrm{ren}} 
=\tilde G_{ijkk}^{\mathrm{ren}} = 0$, and we may write, in general, that
\be
\tilde G_{ijkl}^{\mathrm{ren}} = D_{ijkl} -\frac{1}{3}(x+y) \delta_{ij} \delta_{kl} \ , \
\ee
where 
\be
D_{ijkl} = x \delta_{ik} \delta_{jl} + y \delta_{il} \delta_{jk} \ , \
\ee
with $x$ and $y$ being two independent arbitrary parameters. A straightforward computation of the noise renormalization diagram, Fig. \ref{fig:oneloop}a,
gives us
\begin{equation}
 \tilde C^A_{ijkl} = \frac{g^4}{8} \left( 6 \delta_{ik} \delta_{jl}
 - \frac{1}{4} \delta_{il} \delta_{jk} - \frac{23}{12} \delta_{ij} \delta_{kl} \right) \mbox{,}
\end{equation}
and, as a consequence,
\begin{equation} \label{eq:renormalization_xy}
x = 2 + \frac{3}{2} g^2 \ , \  y = -\frac12 - \frac{1}{16} g^2 \mbox{.}
\end{equation}

Recalling, now, the saddle-point Eqs. (\ref{eq:EOM}) to solve for $\hat \A^{sp}$ in terms of $\A^{sp}$, it turns out that the MSRJD effective action can be 
rewritten in a more compact way, up to the same order in perturbation expansion, as a scalar functional uniquely
dependent on the velocity gradient tensor field $\A(t)$, namely,
\begin{subequations}
\begin{align}
&\Gamma [\A] = \frac{1}{2 g^2} \int_{-\beta}^0 d t \ [ \tilde L^{\mathrm{ren}}_{ij} (\A) D^{-1}_{ijkl} \tilde L^{\mathrm{ren}}_{kl} (\A) ] \ , \  \label{eq:shortGamma}\\
&= \frac{a}{2 g^2} \int_{-\beta}^0 d t \ \mathrm{Tr} [( L^{\mathrm{ren}} (\A))^\T L^{\mathrm{ren}}(\A)] +
 \frac{b}{2 g^2} \int_{-\beta}^0 d t \ \mathrm{Tr} [L^{\mathrm{ren}}(\A) L^{\mathrm{ren}}(\A)] \ , \ \label{eq:longGamma}
\end{align}
\end{subequations}
where
\be
D^{-1}_{ijkl} \equiv a \delta_{ik} \delta_{jl} + b \delta_{il} \delta_{jk} \ , \
\ee
with
\begin{equation} \label{eq:ab_from_xy}
    a = - \frac{x}{y^2 - x^2} \ , \ b = \frac{y}{y^2 - x^2} \ , \     
\end{equation}
and
\begin{equation} \label{eq:L_renormalized}
 L^{\mathrm{ren}}(\A) = L(\A) + \frac{g^2}{16} (4 \A^\T - \A) = \dot \A - V(\A) + \frac{g^2}{16} (4 \A^\T - \A) \mbox{,}
\end{equation}
as it follows from the evaluation of the propagator renormalization diagram, Fig. \ref{fig:oneloop}b.
A clear advantage of the formulation (\ref{eq:longGamma}) is that we do not need to work anymore with a coupled set of saddle-point equations like (\ref{eq:EOM}). Actually, taking into account (\ref{eq:longGamma}), it is only necessary to consider the single saddle-point equation,
\begin{equation} \label{eq:euler-lagrange-gamma}
 \left. \frac{\delta \Gamma[\A]}{\delta A_{ij}} \right|_{\A=\A^{sp}} = 0 \ , \ 
\end{equation}
in order to find the instanton $\A^{sp}(t)$.

It is worth mentioning, in passing, that (\ref{eq:longGamma}) can be alternatively seen as an effective Onsager-Machlup action functional \cite{onsager1953fluctuations} for the RFD model.
\vspace{0.3cm}

\subsection{Instanton Configurations}

\begin{figure}[htb]
 \centering
 \includegraphics{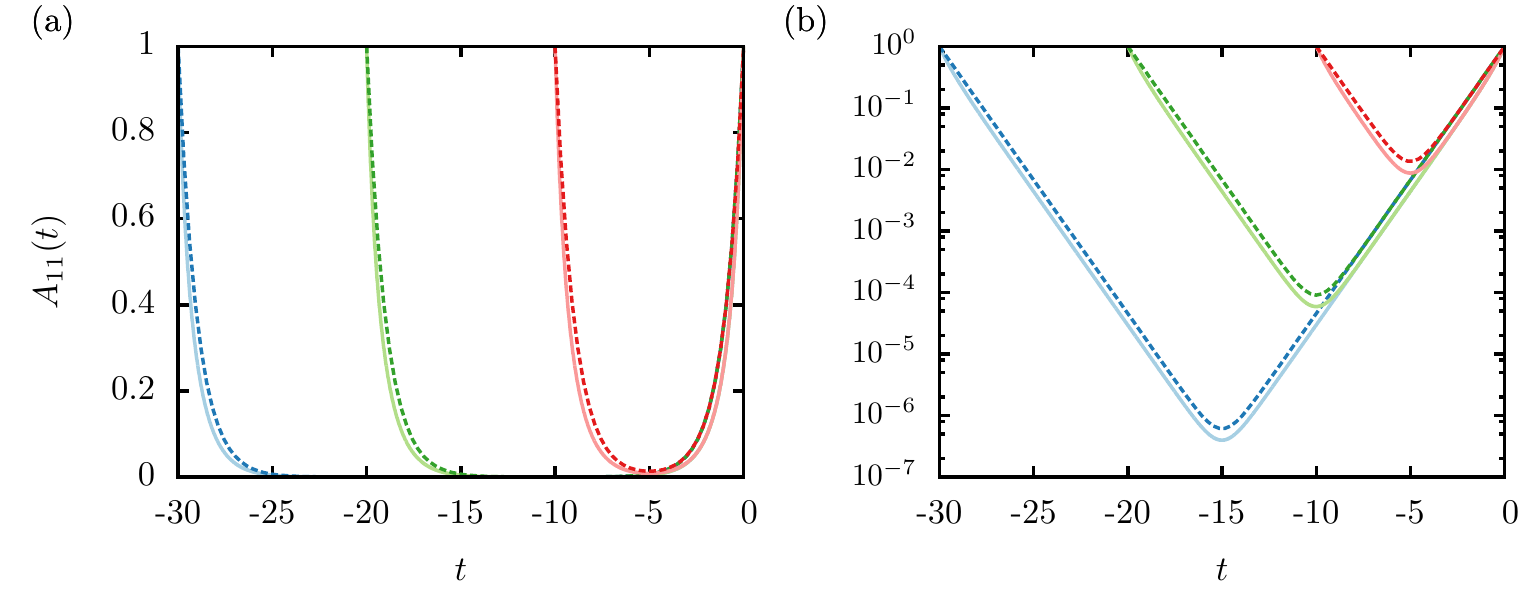}
 \caption{Comparison between approximate (dashed lines) and numerical instantons (solid lines), obtained, respectively from Eq. (\ref{eq:analytical_instanton}) and from the application of the Chernykh-Stepanov method as discussed in Ref. \cite{grigorio2017instantons}, for $c=1$ [that is $\bar A_{11} = 1$, see Eq. (\ref{c_boundary})], $\tau=0.1$, and $g=0.8$ in (a) linear and (b) monolog scales.
 Notice that the approximate and the numerical
 instantons both refer to the RFD model.
 Blue, green and red curves (left to right) correspond to $\beta =30$, $20$ and $10$.
 }
 \label{fig:instanton}
\end{figure}

As emphasized in Sec. II, the RFD model describes the dynamics of incipient turbulent fluctuations at low Reynolds numbers. This fact suggests that the probability measure defined from $\rho (\bar \A)$ is not that far (in some functional sense) from a multivariate Gaussian distribution, even though heavy tails for the marginal PDFs of velocity gradient components can be clearly identified from numerical simulations of Eq. (\ref{eq:dotA}) \cite{ChevPRL}. We are motivated, thus, to devise approximate saddle-point solutions of (\ref{eq:gamma_onsager}) by working with the quadratic truncation of the renormalized effective action, that is,
\begin{equation} \label{eq:gamma_onsager}
 \Gamma_0[\A] \equiv \frac{a}{2 g^2} \int_{-\beta}^0 d t \ \mathrm{Tr}
 \left[\dot \A^\T \dot \A + \A^\T \A \right] +
 \frac{b}{2 g^2} \int_{-\beta}^0 d t \ \mathrm{Tr} \left[\dot \A^2 + \A^2 \right] \mbox{.}
\end{equation}
We just mean that we are interested to solve the approximate saddle-point equation
\begin{equation}\label{eq:approx_EOM}
 \left. \frac{\delta \Gamma_0[\A]}{\delta A_{ij}} \right|_{\A=\A^{sp}} = 0 \Rightarrow  \ddot \A^{sp} - \A^{sp} = 0
\ ,\ 
\end{equation}
subject to the periodic boundary condition (\ref{eq:boundary}),
instead of the exact solution of the complete, non-linear
Euler-Lagrange Eqs. (\ref{eq:euler-lagrange-gamma}).
Instanton solutions of (\ref{eq:approx_EOM}) have the form
\begin{equation} \label{eq:analytical_instanton}
 \A^{sp} (t) = \bar \A f(\beta, t) \mbox{,}
\end{equation}
where the time-periodic function $f(\beta,t)$, defined for $-\beta \leq t \leq 0$, is given by
\begin{equation}
 f(\beta, t) = 2 \frac{\sinh(\beta/2)}{\sinh(\beta)} \cosh(t + \beta/2) \mbox{.}
\end{equation}
It is clear, additionally, that the vertex functions $V_p(\A)$, being homogeneous functions of de\-gree $p$, lead, according to (\ref{eq:analytical_instanton}), to
\begin{equation}
 V(\A^{sp}(t)) = \sum_{p = 1}^4 V_p (\bar \A) [ f(\beta,t) ]^p \mbox{.}
\end{equation}
Taking, now, the above expression together with (\ref{eq:vpowers}) and (\ref{eq:L_renormalized}), the effective action (\ref{eq:longGamma}) can be evaluated from the following scalar contributions,
\begin{equation}
 \int_{-\beta}^0 dt \ \mathrm{Tr} [ (L^{\mathrm{ren}}(\A^{sp}))^\T L^{\mathrm{ren}}(\A^{sp})] = I_1 (\beta) 
 \mathrm{Tr} [ \bar \A^\T \bar \A ] + \sum_{p=1}^4 \sum_{q=1}^4 I_{p+q}  (\beta) H_{p,q}(\bar \A^\T, \bar \A)
\end{equation}
and
\begin{equation}
 \int_{-\beta}^0 dt \ \mathrm{Tr} [ L^{\mathrm{ren}}(\A^{sp})  L^{\mathrm{ren}}(\A^{sp})] = I_1 (\beta) 
 \mathrm{Tr} [ \bar \A^2  ] + \sum_{p=1}^4 \sum_{q=1}^4 I_{p+q} (\beta) H_{p,q}(\bar \A, \bar \A) \ , \
\end{equation}
where $H_{p,q}(X,Y)$ is a (computable) homogeneous scalar function of degrees $p$ and $q$, related, 
respectively, to the matrix variables $X$ and $Y$, and
\begin{equation}
 I_1 (\beta)  \equiv \int_{-\beta}^0 d t [ \dot f(\beta, t) ]^2 \mbox{,}
 \ I_{p+q} (\beta) \equiv \int_{-\beta}^0 d t [ f(\beta, t) ]^{p+q} \mbox{.}
\end{equation}
At asymptotic times, $\beta \rightarrow \infty$, we define $I_p = \lim_{\beta \rightarrow \infty } I_p (\beta)$
to find
\bea
 &I_1 = I_2 = 1 \mbox{,} \  I_3 = 2/3 \mbox{,} \  I_4 = 1/2 \mbox{,}  \nonumber \\
 &I_5 = 2/5 \mbox{,} \  I_6 = 1/3 \mbox{,} \  I_7 = 2/7 \mbox{,} \  I_8 = 1/4 \mbox{.}
\eea
Assembling all the above pieces together, we write the effective action as
\begin{equation}\label{gamma_A}
\begin{split}
&\Gamma[\A^{sp}] \equiv \Gamma( \bar \A) = \\
&=\frac{aI_1}{2 g^2} \mathrm{Tr} [ \bar \A^\T \bar \A ] + \frac{bI_1}{2 g^2} \mathrm{Tr} [ \bar \A^2  ]
 + \sum_{p=1}^4 \sum_{q=1}^4 \frac{I_{p+q}}{2g^2} [ a H_{p,q}(\bar \A^\T, \bar \A)
 + b H_{p,q}(\bar \A, \bar \A) ] \ . \ 
\end{split}
\end{equation}
The normalized vgPDF can now be readily derived from (\ref{nn_vgPDF}) and (\ref{gamma_A}), therefore, as 
\be
\rho(\bar \A) = {\cal{N}} \exp [ - \Gamma(\bar \A) ] \ . \  \label{n_vgPDF}
\ee

It is interesting to compare our approximate instanton solutions (\ref{eq:analytical_instanton}) with accurate numerical solutions in specific cases. As discussed in Ref. \cite{grigorio2017instantons}, diagonal velocity gradient instantons can be obtained from the application of the Chernykh-Stepanov method \cite{chernykh2001large} in connection with the MSRJD action (\ref{eq:the_MSRJD_action}) for the particular boundary conditions,
\bea 
&&\bar A_{11} = -2 \bar A_{22} = -2 \bar A_{33} \equiv c \ , \ \label{c_boundary} \\
&& \bar A_{ij} = 0 {\hbox{  for $i \neq j$}} \ ,  \ 
\eea
where $c$ is an arbitrary constant. The approximate and the numerical instantons for $c=1$, $\tau=0.1$, and $g=0.8$ are both plotted in Fig. \ref{fig:instanton}, for three different values of the $\beta$ parameter. As it can be clearly noticed, the approximate instantons are uniformly close to the exact ones, with very reasonable agreement.
The use of approximate instantons is justified in our treatment by
two important related points:  (i) we actually consider flow regimes
which have moderate Reynolds numbers, so these saddle-point
solutions are actually close to
the exact ones when (ii) we focus on small fluctuations of the velocity
gradient tensor, which already have incipient non-gaussian features.

Since the effective action contains terms up to order $\bar{A}^8$, one could assume that it would be necessary to use saddle-point solutions which approximate the exact instantons up to higher orders. Our perturbative calculations, however, are not merely based on an expansion in powers of $\bar{A}$. Instead, in the cumulant expansion method, we postulate a hierarchy of perturbative contributions (the cumulants), where terms with different orders of $\bar{A}$ are mixed. As actually highlighted in the previous section, the flow regimes we are interested to understand are not associated to asymptotically large values of $\bar{A}$, so that there is no need to saturate the coefficients of higher powers of $\bar{A}$ with all possible contributions, in order to accurately model the PDFs in the range of velocity gradients we explore.

The use of approximate instantons implies, as a counterpart,
that we carefully take into account the effects of first order fluctuations 
around the MSRJD saddle-point action (that would be otherwise absent
if exact instantons were considered from the start).

At this point is worth of emphasizing that the instanton approach in turbulence has been usually applied to derive the behavior of far PDF's tails that decay faster than any simple exponential. However, in the particular case of the RFD model, one can numerically check that the vgPDFs' far tails are subexponential (a point we will not discuss in detail here, to avoid further digressions), so that our use of instantons is all related to the modeling of non-gaussian deviations of vgPDFs around their central peaks.


\section{Numerical Results} 
\label{sec:Results}

We discuss, now, how the analytical predictions based on the above effective action formalism perform in the statistical modeling of turbulent velocity gradient fluctuations.  

\begin{figure}
 \centering
 \includegraphics{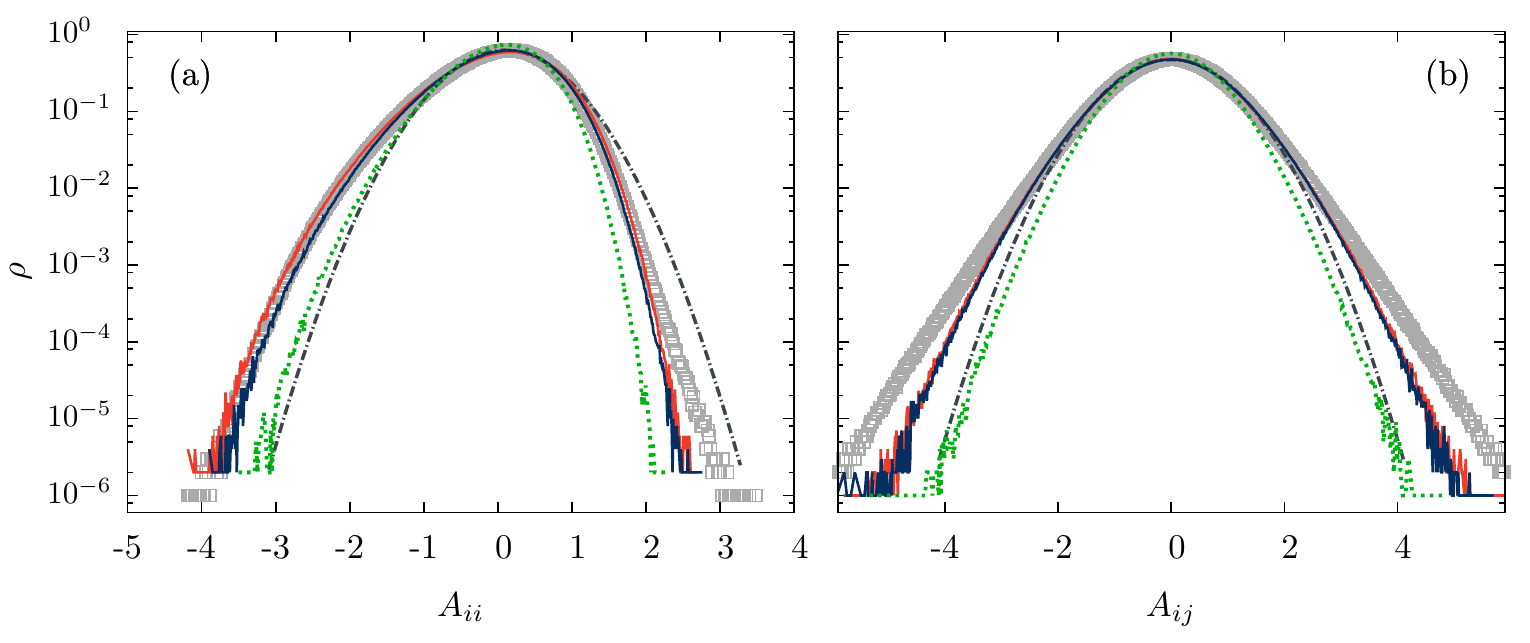}
 \caption{PDFs for the (a) diagonal and (b) off-diagonal velocity gradient components, computed for $\tau=0.1$ and $g=0.8$. Open squares refer to the empirical vgPDFs derived from the numerical
 solutions of the RFD model, Eq.~(\ref{eq:dotA}), while all the other vgPDFs follow from analytical expressions obtained at different improvement levels. Green dashed lines correspond to no effective action renormalization, red (light gray) lines to partial renormalization and blue (dark gray) lines to full renormalization.
 The dotted-dashed gray lines correspond
 to Gaussian fits of the numerical (RFD) data 
 around the vgPDFs' peaks, which clearly show deviations
 from quadratic behavior
 both in the partial and full renormalization schemes.
 The diagonal and off-diagonal empirical vgPDFs have standard deviations and kurtosis given by $\sigma = 0.66$ and $k = 3.3$ and
$\sigma = 0.89$ and $k = 3.7$, respectively.
 }
 \label{fig:pdf-single-g}
\end{figure}

\begin{figure}
 \centering
 \includegraphics[width=\textwidth]{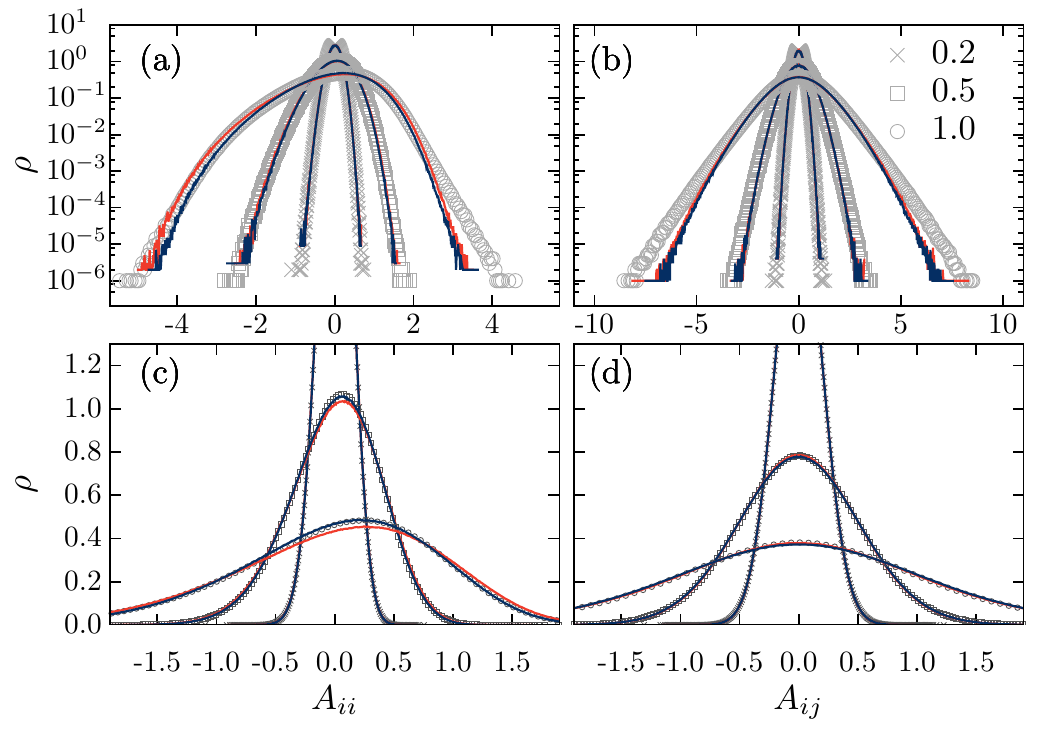}
 \caption{
 PDFs for the diagonal components of the velocity gradient tensor
 at (a) and (c), and its off-diagonal components, at (b) and (d).
 Figures (a) and (b) are in monolog scale, while figures
 (c) and (d) are in linear scale, and they represent the same
 sets of data.
 Symbols refer to the empirical vgPDFs derived from the numerical solutions
 of Eq. (\ref{eq:dotA}), for different values of the random force
 strength $g$, as indicated in the plots; red (light gray) and blue
 (dark gray) lines refer, respectively, to vgPDFs obtained from partial
 and fully renormalized effective actions. For the sake of better
 vizualization, we have not plotted the non-renormalized vgPDFs.}
 \label{fig:pdf-several-g}
\end{figure}

\subsection{Marginal ${\hbox{\textbf{vgPDFs}}}$}\label{ssec:vgpdfs}

Since (\ref{n_vgPDF}) is a multivariate PDF defined on the domain of nine velocity gradient components, there is no way to evaluated it numerically from a multidimensional histogram. Partial analytical integrations which could yield marginal PDFs of $A_{ij}$ are not feasible as well, since $\Gamma [ \bar \A ]$ is not quadratic. We have to resort, in this way, to the analysis of numerical statistical ensembles generated from the vgPDFs given by (\ref{n_vgPDF}). They can be produced along the lines of the Monte Carlo procedure put forward in \cite{MPG}, where random fluctuations of $A_{ij}$ are parametrized by an overcomplete basis of $3 \times 3$ traceless matrices.

Our Monte Carlo samples consisted of sets of $ 8 \times 10^6$ velocity gradient tensors, from which we extracted ensembles of $ 24 \times 10^6$ and $ 48 \times 10^6$ diagonal and off-diagonal velocity gradient components, respectively. An illustrative case for the marginal PDFs of the diagonal and off-diagonal components of the velocity gradient tensor is given in Fig. \ref{fig:pdf-single-g}, for controlling parameters $\tau=0.1$ and $g=0.8$.
For this value of $\tau$, the RFD model leads, as indicated from numerical experiments,
to statistical results similar to the ones observed in
realistic turbulent flows \cite{ChevPRL}
We compare results for four distinct situations: vgPDFs obtained from

\begin{enumerate}[label=(\roman*)]
\item \label{it:DNS} the straightforward numerical simulations of the RFD model, Eq. (\ref{eq:dotA}), with samples containing $10^9$ velocity gradient tensors (which correspond to $2 \times 10^5$  integral times scales), 
\item \label{it:NonR} the saddle-point MSRJD action with no renormalization contributions,
Eq. (\ref{eq:the_MSRJD_action}). 
\item \label{it:NoiseR} the {\it{ partial renormalization}} of the effective MSRJD action, which is renormalized only by the noise contribution,
as given by Eqs. (\ref{eq:effective_with_integrals}) and
(\ref{eq:noise_renormalized}), with $C^B_{ij}=0$ prescribed in
(\ref{eq:potential_renormalized}), and
\item \label{it:FullR} the {\it{full renormalization}} of the effective action, which is renormalized by both the noise and the propagator contributions,
as given by Eqs. (\ref{eq:effective_with_integrals}),
(\ref{eq:noise_renormalized}), and (\ref{eq:potential_renormalized}).
\end{enumerate}

As it can be clearly seen from Fig. \ref{fig:pdf-single-g}, a great improvement is attained with the use of renormalized actions. Noise renormalization is found to be the leading contribution, and for this reason we may refer to the propagator renormalization contribution as the subleading one.

When we compare the results from partial and full renormalization schemes, still focusing on Fig. \ref{fig:pdf-single-g}, it seems that they would be essentially equivalent, with small differences observed, at first sight, mainly for the PDF tails of diagonal velocity gradient components. Actually, we should not be misled by visual inspection. As we will show, the core regions of these distributions, which already depart from Gaussian behavior, are much more accurately described by the vgPDFs obtained through fully renormalized effective actions.

One may wonder how the modeled vgPDFs plotted in Fig. \ref{fig:pdf-single-g} (the green, red and blue lines) would change if exact (numerical) instantons \cite{grigorio2017instantons}, were used instead of the approximated (but analytical) ones considered in this work. It is clear that when renormalization is absent, the use of exact instantons leads in general to better and reasonable vgPDFs for small values of $g$. For larger values of $g$ ($g > 0.4$, roughly), fluctuations become more important and have to be necessarily taken into account for a proper modeling of the vgPDFs \cite{grigorio2017instantons}.

A more extensive comparison between vgPDFs is provided in Fig. \ref{fig:pdf-several-g}, where we examine cases up to the border line for the application of perturbation theory, which takes place around $g \simeq 1$.
This upper limit
can be estimated from the perturbative corrections (\ref{eq:renormalization_xy})
and appreciated from the results of Fig. \ref{fig:pdf-several-g}:
above $g=1.0$, the renormalized vgPDFs are noted to deviate
in a more expressive way from the numerical ones.
It is to be remarked here that the definition of an upper bound for the coupling
constant $g$ is by no means a sufficient condition for the 
validity of the perturbative expansion,
since it is important that both $g$ and $A$
are not too large for the consistency of the cumulant expansion
approach.
There is a clear interplay between these quantities, 
since the variance of $A$ scales
as $g^2$ for small values of $g$, though this yields
interesting information only around the vgPDFs' peaks. 
We emphasize, for the sake of clarity, that all the numerical results shown in Figs. \ref{fig:pdf-single-g}
and \ref{fig:pdf-several-g}, as well as all other figures,
refer to the RFD model, Eq. (\ref{eq:dotA}). It is also worth noting that the corrections provided by the full renormalization scheme can be clearly appreciated for the off-diagonal vgPDFs, plotted in linear scales, as in Fig. \ref{fig:pdf-several-g}c.

The standard deviations and kurtoses of the several investigated vgPDFs are reported in Table \ref{tab}. We verify, thus, that the renormalization procedures lead in general to vgPDFs which closely fit the empirical ones by several standard deviations around their peaks, within the validity range of the perturbative expansion.
The modeled standard deviations agree reasonably well with the ones 
evaluated through the numerical simulations of the RFD stochastic equations.
The comparison between kurtoses, on the other hand, has some deviations
which are mainly due to the slow decay of the far 
tails of the vgPDFs, which are out of the reach of the present approach.

To analyze the
observed ranges of agreement between the fully renormalized and the empirical
vgPDFs in a more quantitative way, we define them as the velocity gradient regions where the calculated 
perturbative corrections correspond to a given fraction, say $20\%$, of the saddle-point action.
The values of $\bar A$ that are obtained from this prescription establish
an estimate for the border of validity of perturbation theory,
and turn out to be well described by an approximate power-law relation,
$ \bar A/g \approx 1.29 \ g^{-0.41} $. Following such a procedure, we find that the limit of 
applicability of perturbation theory along the lines of the cumulant expansion is actually 
compatible with the qualitative arguments addressed before.

\begin{table}
\centering
\begin{tabular}{|c|c|c|c|c|c|c||c|c|c|c|c|c|c|}
\hline
\textbf{g}  & \multicolumn{2}{c|}{0.2} & \multicolumn{2}{c|}{0.5} & \multicolumn{2}{c||}{1.0} 
& \multicolumn{2}{c|}{0.2} & \multicolumn{2}{c|}{0.5} & \multicolumn{2}{c|}{1.0}
\\ \hline
 {\textbf{Statistical Ensembles}} & D & OD & D & OD & D & OD & D & OD & D & OD & D & OD \\ \hline
Numerical RFD & 0.14 & 0.20 & 0.39 & 0.52 & 0.86 & 1.15 & 3.05 & 3.03 & 3.25 & 3.23 & 3.23 & 3.87 \\ \hline
No Renormalization & 0.14 & 0.20 & 0.35 & 0.48 & 0.68 & 0.88 & 3.03 & 3.01 & 3.12 & 3.09 & 3.19 & 3.33 \\ \hline
Partial Renormalization & 0.14 & 0.20 & 0.39 & 0.52 & 0.90 & 1.13 & 3.03 & 3.01 & 3.14 & 3.12 & 3.14 & 3.51 \\ \hline
Full Renormalization &  0.14 & 0.20 & 0.39 & 0.52 & 0.84 & 1.13 & 3.03 & 3.01 & 3.14 & 3.10 & 3.14 & 3.39 \\ \hline

 \multicolumn{1}{c}{} & \multicolumn{6}{c}{ $ \vdash$ 
 \textbf{~Standard Deviations} $\dashv$} & 
 \multicolumn{6}{c}{ $ \vdash$ \textbf{~~~~~~~ Kurtoses ~~~~~~~}  $\dashv$ } \\ 

\end{tabular}
\caption{Standard deviations and kurtoses associated to the vgPDFs shown in Fig. \ref{fig:pdf-several-g}. The labels D and OD stand for statistical ensembles of diagonal and off-diagonal velocity gradient components, respectively. These ensembles are characterized, besides the D/OD classification, from specifying how their associated vgPDFs are obtained,
according to four alternative schemes, described as the items
\ref{it:DNS} to \ref{it:FullR} in Subsec. \ref{ssec:vgpdfs}:
numerical simulations of the RFD model, non-renormalized saddle-point MSRJD actions,
partially renormalized effective actions, and fully renormalized effective actions.
}
\label{tab}
\end{table}

\subsection{Joint Statistics of the Velocity Gradient Invariants {\hbox{$Q$}} and {\hbox{$R$}} }

\begin{figure}[htb]
 \centering
 \includegraphics[width=\textwidth]{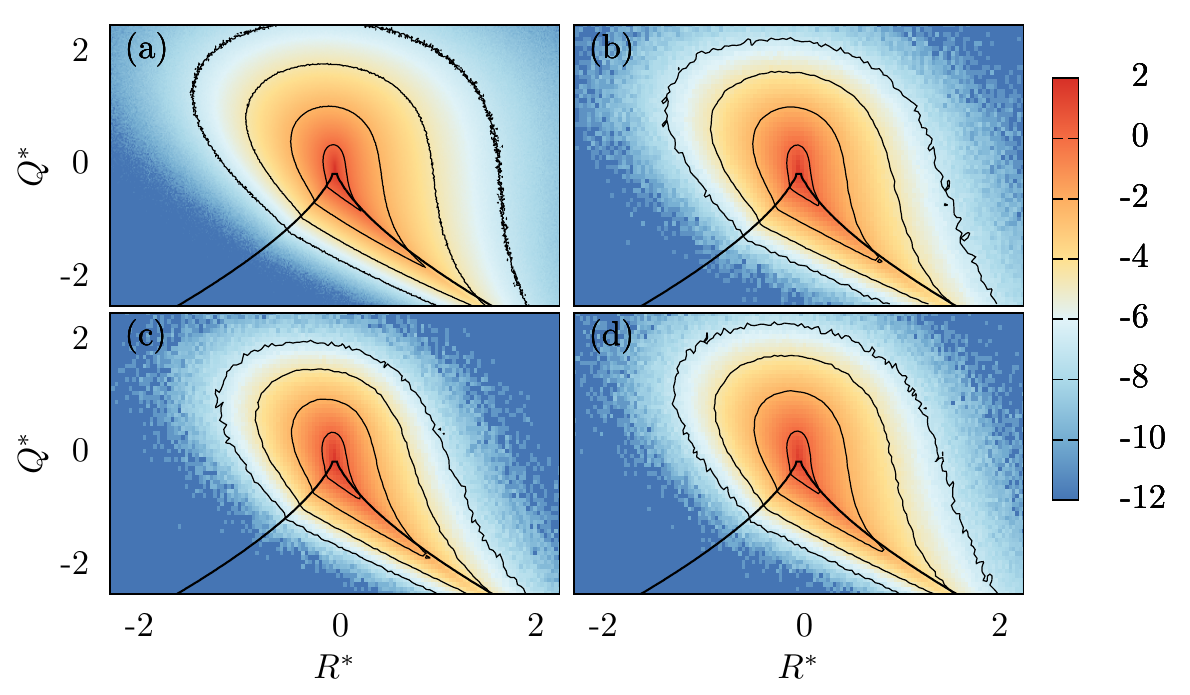}
 \caption{Joint PDFs (and their level curves) of the velocity gradient invariants $Q^*$ and $R^*$, as obtained from (a) numerical simulations of the RFD model,
 (b) the analytical approach based on the non-renormalized effective MSRJD action,
 (c) the noise renormalized effective action and (d) the fully renormalized
 effective action.
 Solid lines represent the ``Vieillefosse line'', defined by $(27/4)R^2 + Q^3 = 0$ (a constraint which holds for the inviscid evolution of the velocity gradient tensor). The data correspond to the RFD parameters $\tau=0.1$ and $g=0.8$,
 and the color bar scale corresponds to powers of 10
 in the visualization of the joint probability distribution functions.
 }
 \label{fig:QR}
\end{figure}

The pair of velocity gradient invariants $Q \equiv - \mathrm{Tr}(\A^2)/2$ and $R \equiv -\mathrm{Tr}(\A^3)/3$ have been extensively used in the recent literature as important observables for the investigation of structural aspects of turbulence \cite{TsinoberInformal}. Turbulent flow regions can be dominated by enstrophy ($Q>0$) or strain ($Q<0$) and, independently, by compression ($R>0$) or stretching ($R<0$) dynamics. It is interesting to work with a dimensionless version of these invariants, 
\begin{equation}
 Q^* = - \frac{\mathrm{Tr}(\A^2)}{2 \langle \mathrm{Tr}(S^2) \rangle } \ \ \mbox{and} \ \ 
 R^* = - \frac{\mathrm{Tr}(\A^3)}{3 \langle \mathrm{Tr}(S^2)  \rangle^{3/2} } \mbox{,}
\end{equation}
where $S = (\A + \A^\T)/2$ is the usual strain rate tensor, the symmetric part of the velocity-gradient tensor. The joint PDF of $Q^*$ and $R^*$ shows a characteristic teardrop shape, as observed from direct numerical simulations of turbulence \cite{leorat1975turbulence,Vieillefosse1982,Vieillefosse1984}, and is qualitatively well reproduced by the RFD model \cite{ChevPRL}. Relying on Monte Carlo ensembles,
in the same fashion as in the previous discussion on marginal vgPDFs, we find that the joint PDFs of $Q^*$ and $R^*$ derived from the full renormalized effective MSRJD action, are in good agreement with the ones obtained from the numerical simulation of Eq. (\ref{eq:dotA}),
as can be seen from the example given in Fig. \ref{fig:QR}.

\subsection{Local Stretching Exponents}

Comparisons between predicted and empirical PDFs can be sometimes a delicate issue, since both of them have to be normalized to unit, and pointwise matching can be lost, even if asymptotic expressions for their tails or cores are correctly derived from theoretical analyses. In order to test the relevance of modeling approaches, as an alternative to simple PDF fitting, one may rely on the concept of local stretching exponents, described as follows, for a non-specific PDF $\rho = \rho(\xi)$ of some real random variable $\xi$. Assuming that this distribution 
belongs to the large class of PDFs that can be written as
\be
\rho(\xi) = \rho_0 \exp[ - S(\xi) ] \ , \
\ee
where $\rho_0$ is just a normalization constant and $S(\xi)$ is a non-negative and monotonically increasing function of $|\xi|$, we define, then, the PDF local stretching exponent $\theta(\xi)$ as
\be
\theta(\xi) = \frac{d \ln S (\xi)}{d \ln |\xi|}  \ . \  \label{theta-xi}
\ee
It is clear, from Eq. (\ref{theta-xi}), that $S(\xi) \sim |\xi|^{\theta(\xi)}$ will hold in a local sense, if $\theta(\xi)$ is a slowly varying function of $\xi$.

Local stretching exponents of velocity gradient PDFs have been previously investigated in the context of Burgers turbulence at high Reynolds numbers \cite{grafke2015relevance}. Taking $\xi = \partial_x u$ to be the spatial velocity derivative of the Burgers one-dimensional velocity field, and $S(\xi)$ to be the saddle-point MSRJD action evaluated in the domain of viscous instantons \cite{balkovsky1997intermittency}, empirical and predicted values of $\theta(\xi)$ were then compared in Ref. \cite{grafke2015relevance} for the PDF tails that describe large negative fluctuations of $\xi$.

It is interesting to address a similar discussion in the RFD model, where $\xi$ can be taken
to be a diagonal or a off-diagonal component of the velocity gradient tensor. It is important, however, to note that the discussion of Ref. \cite{grafke2015relevance},
which focuses on the far tails of PDFs, 
where fluctuations around the instanton become less relevant (and the perturbative
analysis of fluctuations via the cumulant expansion method would break down).
In this work, on the other hand, our attention is centered on the perturbative non-Gaussian deviations of the vgPDF shapes at the onset of turbulence.
This means, as a counterpart, that a straightforward application of Eq. (\ref{theta-xi})
to determine the local stretching exponent of the vgPDFs for diagonal and off-diagonal
velocity gradients would be problematic for small $\xi$,
since both the saddle-point and renormalized MSRJD actions vanish for vanishing velocity gradients.

\begin{figure}[htb]
 \centering
 \includegraphics{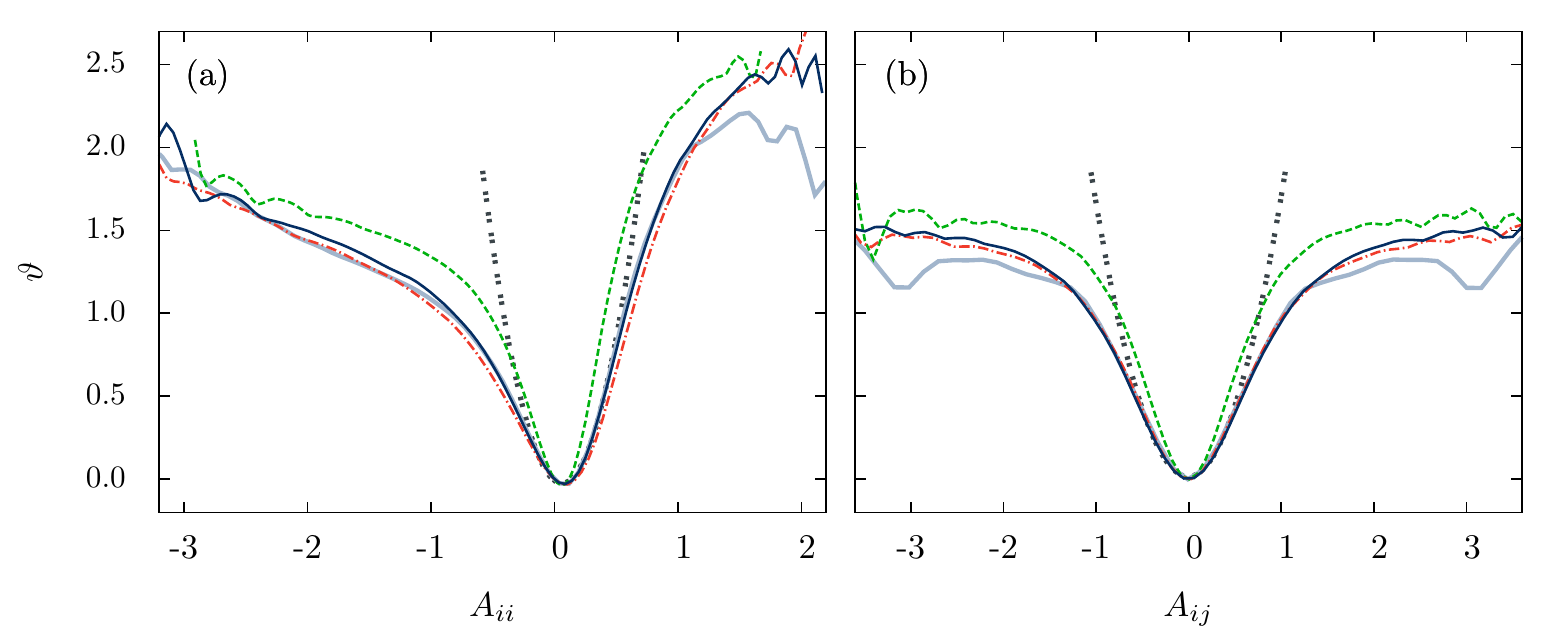}
 \caption{Modified local stretching exponents for the vgPDFs of (a) diagonal and (b) off-diagonal velocity gradient components, evaluated for the RFD model with controlling parameters $\tau=0.1$ and $g=0.8$.
 Stretching exponents are evaluated for vgPDFs derived from (i) the numerical solution of the RFD stochastic equations (gray lines), (ii) the saddle-point MSRJD action (green dashed  lines), (iii) partially renormalized effective MSRJD actions (red dashed-dotted lines), and (iv) fully renormalized effective MSRJD actions (dark blue lines).
 The black dotted lines are parabolas which would give the
 modified stretched exponents if the vgPDFs were exactly gaussian.
 }
 \label{fig:exponent}
\end{figure}

In order to bypass these difficulties, we introduce the modified local stretching exponent $\vartheta(\xi)$ as follows
\be
\vartheta(\xi) = -\frac{d \ln \rho (\xi)}{d \ln |\xi|}  \ . \  \label{theta-xi2}
\ee
Considering that within the vgPDF cores $\theta(\xi)$ is slowly varying, we get
\be
\vartheta(\xi) \simeq \theta(\xi) S(\xi) = \theta(\xi) \ln \left ( \frac{\rho_0}{\rho(\xi)} \right )
\ , \ \label{theta-xi3}
\ee
where, above, $\xi$ represents an arbitrary component of the velocity gradient tensor.
The exponent $\vartheta(\xi)$ would be proportional to $\xi^2$ for a Gaussian distribution,
but we can see clear deviations from a parabolic behavior for the exponents
plotted in Fig. \ref{fig:exponent}.
Eq. (\ref{theta-xi2}) can be used, therefore, to produce rigorous validity tests of the instanton approach to the RFD model, since $\vartheta(\xi)$ depends, as particularly indicated by Eq. (\ref{theta-xi3}), on different defining aspects of the probability distribution $\rho(\xi)$.

Paying careful attention to the robustness of results, we have used high-order B-splines to interpolate the marginal vgPDFs, in such a way that it is not necessary to worry with numerical errors that could be associated to the derivative operation in Eq. (\ref{theta-xi2}). We recall, now, that the pointwise error in determining a general PDF $\rho(\xi)$ from uncorrelated numerical data is
\be
\sigma_\rho (\xi) = \sqrt{\frac{\rho(\xi)(1-\rho(\xi) \delta)}{N \delta}} 
\simeq \sqrt{\frac{\rho(\xi)}{N \delta}}  \ , \ \label{sigma-rho}
\ee
where $\delta$ is the bin size and $N$ is the number of elements in the numerical samples. The meaning of (\ref{sigma-rho}), when working with smooth interpolations, like the ones given by B-splines, is that the estimated PDF can be written, in principle, as
\be
\rho(\xi) = \bar \rho(\xi) + \phi(\xi) \sigma_{\bar \rho(\xi)} \ , \ \label{phi}
\ee
where $\bar \rho(\xi)$ denotes the exact (unknown) PDF, and the modulating function $\phi(\xi)$, with $|\phi(\xi)| < 1$, is assumed to be as smooth as 
$\sigma_{\bar \rho(\xi)}$, that is $|\phi'/\phi | \simeq
| \sigma_{\bar \rho(\xi)}'/ \sigma_{\bar \rho(\xi)} | =
| \bar \rho'/ \bar \rho |/2 $.

From Eqs. (\ref{theta-xi2}), (\ref{sigma-rho}), and (\ref{phi}), we find, thus, that the propagated uncertainty in the evaluation of $\vartheta(\xi)$ has the upper bound
\be
\sigma_\vartheta (\xi) = \vartheta(\xi) \frac{3 \sigma_\rho (\xi)}{4 \rho(\xi)} = \vartheta(\xi) \frac{3}{4\sqrt{N \delta \rho(\xi)}} 
\ . \
\ee
Taking $N \geq 24 \times 10^6$, $\delta \simeq 0.1$, and $\rho(\xi) > 0.1$ within the vgPDFs cores (see Fig. \ref{fig:pdf-several-g}), 
it follows that \
\be
\frac{\sigma_\vartheta (\xi)}{ \vartheta(\xi)} < 1.5 \times 10^{-3}  \ , \   \label{sigma-vartheta}
\ee
so that the modified local stretching exponents are in fact determined with excellent precision in the velocity gradient domains of interest.

As it can be seen from Fig. \ref{fig:exponent}, while both partially or fully renormalized effective MSRJD actions lead to equivalent accurate predictions for the stretching exponents $\vartheta(\xi)$ of the vgPDFs of off-diagonal velocity gradient components, the diagonal case is only accurately modeled by the full renormalization scheme in the range $|A_{ii}| < 1$. Partial renormalization yields, for the diagonal case, predictions missing the empirical values of $\vartheta(\xi)$ in the vgPDF core region by systematic errors of the order of $5\%$, which, according to (\ref{sigma-vartheta}), are considerably greater than the standard deviations of measurement precision.

As the velocity gradients increase in absolute value, we note, from Fig. \ref{fig:exponent}, that the accurate agreement between the predicted and the empirical stretching exponents ends in a somehow abrupt way, even before their evaluations become unreliable. The main reason underlying this phenomenon is the unavoidable breakdown of the perturbative expansion around instantons, when cumulants of higher orders cannot be neglected anymore in comparison with the second order contributions (\ref{eq:diag_a_effective})  and (\ref{eq:diag_b_effective}).


\section{Conclusions} \label{sec:Conclusions}

This work provides a detailed report on the field theoretical approach to the problem of Lagrangian intermittency,
as described along the lines of the RFD model \cite{ChevPRL}. Motivated by the promising results advanced in 
Ref. \cite{MPG}, we found that the analytical expressions for the vgPDFs can be further improved from the consideration 
of additional contributions to the renormalized MSRJD effective action, which rely, ultimately, on the cumulant integration
of fluctuations around instanton configurations.

It is important to emphasize that the present formalism - not restricted at all
to the specific case of the RFD model
-- yields accurate results for the core regions (defined within a few standard
deviations around central peaks) of vgPDFs, 
at the onset of turbulence, when fat tails start to show up. This is due to
the fact that under such conditions the usual cumulant expansion method is 
a reliable perturbation technique.
At the far PDF tails, on the 
other hand, the cumulant expansion breaks down, but, as a counterpart,
the role of fluctuations
around the instantons is assumed to be
less relevant than the one at the PDF cores.

We accentuate that we have not used exact instantons in our study of the RFD model,
but rather a linear approximation to the instanton equations, which,
for the range of Reynolds numbers of interest, are found to be meaningful, after
a perturbative treatment of fluctuations is carried out, in order to 
reproduce the shape of vgPDFs around their non-gaussian cores.

It would be interesting to investigate, in connection with the far vgPDFs' tails, 
possible improvements that would follow from the use of exact instantons and the
integration of fluctuations around them by means of the standard path-integal WKB approach \cite{wkb}. 
We should have in mind, however, that exact instantons can be computed only in a numerical way, 
with the help of the Chernykh-Stepanov procedure \cite{chernykh2001large}, as it became clear from previous
studies of the Burgers problem \cite{grafke2015instanton} and the RFD model \cite{grigorio2017instantons}.

A particularly interesting aspect in having analytical expressions for vgPDFs, as the ones we have obtained, is that they are joint PDFs, and, in this way, may be used to generate Monte Carlo ensembles for the analysis of conditioned statistics phenomena (like the alignment 
correlations between vorticity and the strain principal axes \cite{TsinoberInformal}), which can be considerably larger than the 
corresponding ensembles produced from the straightforward numerical solution of modeling stochastic equations.

Extensions of the analytical methodology discussed in this work to other stochastic hydrodynamic systems, as the Burgers model or 
three-dimensional turbulence, as well as the recent improved variations of the RFD model \cite{johnson2016closure,pereira2018multifractal}, offer no 
major conceptual difficulties and are deserved for future research.

\section*{Acknowledments} \label{sec:Acknowledments}

G.B.A. and L.M. thank CNPq for financial support.
R.M.P. thanks CAPES and FACEPE for financial support and L. S. Grigorio for interesting discussions.
G.B.A. also thanks M. Moriconi for useful discussions.

\bibliographystyle{apsrev4-1}
\bibliography{rfd}

\begin{thebibliography}{49}
\bibitem [{com()}]{comment1}%
  \BibitemOpen
  \href@noop {} {}\bibinfo {note} {We have in mind the large class of additive
  noise models, but our arguments can be straightforwardly extended to the
  cases of stochastic differential equations with multiplicative
  noise.}\BibitemShut {Stop}%
\end{thebibliography}

\end{document}